\documentclass[aps,prd,11pt]{revtex4}
\usepackage{graphicx}
\usepackage{color}
\newcommand{\half}{\frac{1}{2}}
\newcommand{\tr}{\mbox{tr}}

\newcommand{\eq}[1]{eq.~\ref{#1}}

\newcommand{\be}{\begin{equation}}
\newcommand{\ee}{\end{equation}}
\newcommand{\bea}{\begin{eqnarray}}
\newcommand{\eea}{\end{eqnarray}}
\newcommand{\azAngle}{\varphi}
\begin{document}
\begin{center}
{\large \bf  2+1 Dimensional Georgi-Glashow Instantons in Weyl Gauge}
\end{center}
\centerline{ Gerald V.~Dunne and Vishesh Khemani }
\begin{center}
Department of Physics\\
University of Connecticut\\
Storrs, CT 06269-3046
\end{center}
\smallskip
\centerline{\large\bf Abstract}
\bigskip
{\small 
Semiclassical instanton solutions in the 3D SU(2) Georgi-Glashow model are transformed into the Weyl gauge. This illustrates the tunneling interpretation of these instantons and provides a smooth regularization of the singular unitary gauge. The 3D Georgi-Glashow model has both instanton and sphaleron solutions, in contrast to 3D Yang-Mills theory which has neither, and 4D Yang-Mills theory which has instantons but no sphaleron, and 4D electroweak theory which has a sphaleron but no instantons. We also discuss the spectral flow picture of fundamental fermions in a Georgi-Glashow instanton background.
}

\bigskip

\leftline{\it \small Keywords:~\parbox[t]{15cm}{
monopole, instanton, sphaleron, vortex}}

\leftline{\it \small PACS:~\parbox[t]{15cm}{11.10.Kk, 11.15.Kc}}
%
\bigskip
\section{Introduction} 

A detailed understanding of confinement remains one of the great challenges of quantum chromodynamics (QCD). The 3 dimensional Georgi-Glashow (GG) model is an interesting model system for studying confinement, as it exhibits confinement at zero temperature due to instanton effects \cite{polyakovgg}. Confinement arises in this system due to interactions of a plasma of instantons. On the other hand, while the physical role of instantons as semiclassical tunneling solutions is well understood in 4D Yang-Mills theory \cite{jackiw-rebbi,callan-dashen-gross1,roman,schafer,shifman,falk,diakonov}, this has not been spelled out in detail for the 3D GG model. Further motivation for this study comes from the fact that the deconfinement phase transition in the 3D SU(2) GG model has recently been found analytically \cite{dkkt,kogan-kovner}, and is in the ${\bf Z}_2$ (Ising) universality class, in agreement with general arguments of Svetitsky  and Yaffe \cite{svetitsky-yaffe} relating global symmetries and the universality classes of phase transitions. 
This correspondence between the global ${\bf Z}_2$ symmetry and the universality class also has important implications for abelian projections \cite{ogilvie}. The analysis of \cite{dkkt} used dimensional reduction (see also \cite{agasian}) and an interplay between monopole (instanton) and W-boson densities. The results of \cite{dkkt} have recently been confirmed by lattice simulations \cite{barresi}, and corrections to the BPS estimate of the transition temperature have been computed in \cite{kovchegov-son}. There is also important work on the relation between semiclassical and nonperturbative lattice treatments \cite{laine,philipsen,hart,davis}, which suggests the existence of an analytic connection between the Higgs and confining phases. In this paper we present a pedagogical discussion of the tunneling interpretation of instantons in the 3D GG model. Mathematically, these GG instantons are just 't Hooft-Polyakov monopoles \cite{thooft,polyakov,manton}, and we show how to transform such classical solutions into the Weyl ($A_3=0$) gauge, which is an unfamiliar gauge from the monopole viewpoint, but very natural from the 3D GG viewpoint.

\section{The 3D Georgi-Glashow Model}
\label{3dgg}

Consider the $SU(2)$ Georgi-Glashow (GG) model in 2+1 dimensions. This model is an $SU(2)$ Yang-Mills gauge theory minimally coupled to a scalar field, $h$, in the adjoint representation, and with a symmetry breaking quartic scalar potential. The 3D Euclidean space action is
\be
S = \int d^3 x \left[ \half \tr\left( F_{\mu\nu} F_{\mu\nu} \right) + \half D_\mu h^a D_\mu h^a + \frac{\lambda}{4}\left( h^a h^a - v^2 \right)^2  \right] \, .
\label{eq:action}
\ee
Our conventions are
\bea
A_\mu &=& \half A_\mu^a \tau^a \, , \nonumber \\
F_{\mu\nu} &=& \partial_\mu A_\nu - \partial_\nu A_\mu - i g [A_\mu , A_\nu] \, , \nonumber \\
h &=& \half h^a \tau^a \, , \nonumber \\
D_\mu h &=& \partial_\mu h - i g [A_\mu , h] \quad , \quad D_\mu h^a=\partial_\mu h^a+g\, \epsilon_{abc}\, A_\mu^b h^c\, ,
\label{eq:notation}
\eea
where  $\tau^a$ ($a=1, 2, 3$) denotes the $2\times 2$ Pauli matrix generators of $su(2)$. Under an $SU(2)$ gauge transformation, $U$, the fields transform as:
\bea
A_\mu&\to& U A_\mu U^\dag+\frac{i}{g}U \partial_\mu U^\dag \, ,\nonumber\\
h&\to& U\, h\, U^\dag\, .
\label{eq:gaugetrans}
\eea 
Note that in $2+1$ dimensions the coupling $g$ has dimensions of $({\rm mass})^{1/2}$, as do the fields $h$ and $A_\mu$.

Perturbatively, the $SU(2)$ symmetry is spontaneously broken to $U(1)$ when the scalar field develops a vacuum expectation value $v$.  This results in a massless Abelian gauge field (photon), a Higgs field with mass $M_H = \sqrt{2 \lambda}v$, and two massive gauge fields ($W^\pm$) with degenerate mass $M_W = g v$.  Non-perturbative effects from an instanton plasma give a small nonperturbative mass to the photon, and lead to the linear confinement of the massive gauge fields at zero temperature \cite{polyakovgg}. These instantons are solutions to the classical Euclidean equations of motion:
\bea
D_\mu D_\mu h &=& \lambda \left( h^a h^a - v^2 \right) h \, , \nonumber \\
D_\mu F_{\mu\nu} &=& i g [D_\nu h , h] \, .
\label{eq:fullEqnsOfMotion}
\eea
Mathematically, these instanton solutions are identical to the 't Hooft - Polyakov monopoles \cite{thooft,polyakov} of the corresponding $3+1$ dimensional theory, since the Euclidean 3-dimensional action (\ref{eq:action}) is identical to the classical potential energy of the $3+1$ dimensional model. However, {\it physically} their role is of course completely different.

Finite action solutions to the Euclidean classical equations of motion (\ref{eq:fullEqnsOfMotion}) are characterized by an instanton number (monopole number in the $(3+1)$-dimensional context) which can be defined as follows \cite{thooft,arafune}. Due to the potential term in (\ref{eq:action}), at the boundary of space-time the scalar field $h$ must tend to a value satisfying $h^a h^a =v^2$. Thus, such an $h$ defines a map 
\bea
h:S^2_\infty \mapsto S^2_{\rm internal}
\label{eq:s2map}
\eea
from the $S^2_\infty$ boundary of space-time to the internal isospace 2-sphere, $S^2_{\rm internal}$, defined by $h^a h^a =v^2$.
These maps are characterized by an integer degree, which is the instanton number, and which can be expressed in terms of the fields as follows \cite{arafune}. Define 't Hooft's abelian field strength \cite{thooft} away from the zeros of the scalar field as 
\bea
F_{\mu\nu}=\hat{h}^a F_{\mu\nu}^a +\frac{1}{g}  \epsilon_{abc} \hat{h}^a D_\mu \hat{h}^b D_\nu \hat{h}^c 
\label{eq:thooftF}
\eea
where the unit Higgs field, $\hat{h}=\hat{h}^a \frac{\tau^a}{2}$, is defined as
\bea 
\hat{h}^a\equiv \frac{h^a}{\sqrt{h^b h^b}}
\label{eq:unithiggs}
\eea
This can also be written as  \cite{arafune}
\bea
F_{\mu\nu}&=&\left(\partial_\mu B_\nu-\partial_\nu B_\mu\right) +\left(\frac{1}{g} \epsilon_{abc} \hat{h}^a \partial_\mu \hat{h}^b \partial_\nu \hat{h}^c \right)\nonumber\\
&\equiv& M_{\mu\nu}+H_{\mu\nu}
\label{eq:split}
\eea
where the projected abelian gauge field is
\bea 
B_\mu\equiv \hat{h}^a A_\mu^a
\label{eq:abelianfield}
\eea
Then the local instanton charge density is
\bea
k=\frac{g}{2}\epsilon_{\mu\nu\rho} \partial_\mu F_{\nu\rho}
\label{eq:density}
\eea
For configurations without line singularities, $M_{\mu\nu}$ does not contribute (by the Bianchi identity for $B_\mu$) to the local instanton charge density $k$, and so the instanton number can be defined in terms of the scalar fields only \cite{arafune}:
\bea 
I&=& \frac{1}{4\pi} \int d^3 x \, k \nonumber\\
&=& \frac{1}{8\pi} \int d^3 x\, \epsilon_{\mu\nu\rho} \epsilon_{abc} \partial_\mu \left(\hat{h}^a \partial_\nu \hat{h}^b \partial_\rho \hat{h}^c\right) \nonumber\\
&=& \frac{1}{8\pi} \int_{S^2_\infty} (d^2s)_\mu \, \epsilon_{\mu\nu\rho} \epsilon_{abc} \hat{h}^a \partial_\nu \hat{h}^b \partial_\rho \hat{h}^c
\label{eq:instantonnumber}
\eea
The last line clearly shows the topological interpretation of the instanton number $I$ as the degree of the map in (\ref{eq:s2map}). 

The classical Euclidean equations of motion (\ref{eq:fullEqnsOfMotion}) form a set of coupled nonlinear partial differential equations and are difficult to solve. One simplifying approach is to search for solutions within a radially \cite{thooft,polyakov} or axially \cite{rebbi-rossi,kleihaus,shnir} symmetric ansatz for the fields. The only nontrivial instantons described by the spherical ansatz have instanton number $| I |=1$ \cite{weinberg}. In the BPS limit (when the scalar self-coupling $\lambda\to 0$), the equations simplify somewhat and there are more powerful techniques for finding solutions without resorting to ansatze \cite{manton}. 

\subsection{Spherical Ansatz for Single Instanton}
\label{spherical}

The simplest GG instanton is described by the spherical ansatz \cite{thooft,polyakov}:
\bea
h^a(x) &=& v f_h(r) \frac{x^a}{r} \, , \nonumber \\
A_\mu^a (x) &=& \frac{f_A(r)}{g r} \epsilon_{a \mu b} \frac{x^b}{r} \, .
\label{eq:spherical}
\eea
Here $r=\sqrt{x_\mu x_\mu}$ is the 3D radial Euclidean space-time coordinate.  With this ansatz, the classical Euclidean equations of motion (\ref{eq:fullEqnsOfMotion}) reduce to the following coupled ordinary differential equations for the two profile functions:
\bea
\frac{d^2 f_h}{d r^2} + \frac{2}{r}\frac{d f_h}{dr} &=& \frac{2 f_h}{r^2}(1-f_A)^2 + \lambda v^2 f_h (f_h^2 - 1) \, , \nonumber \\
\frac{d^2 f_A}{d r^2} &=& \frac{f_A}{r^2} (f_A - 1)(f_A - 2) + g^2 v^2 f_h^2 (f_A - 1) \, , \nonumber\\ 
f_h(0)= f_A(0) &=& 0\, , \nonumber\\
f_h(r \rightarrow \infty) = f_A(r \rightarrow \infty) &=& 1\, .
\label{eq:GGode}
\eea
The asymptotic scalar field configuration is a one-to-one winding 1 map from the boundary of space-time ($S^2_{\infty}$) to $S^2_{\rm internal}$, spanned by $h^a/|h|$.  Thus this spherical GG instanton has instanton number 1.
\begin{figure}
\centerline{\includegraphics[width=6cm]{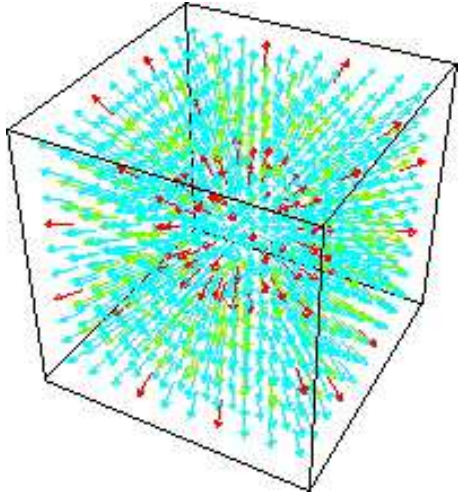}}
\caption{\label{fig1}
The familiar hedgehog form of the scalar field $h^a$ in the spherical ansatz (\protect{\ref{eq:spherical}}) for the single instanton. Contrast with the corresponding plot of $h^a$ in the Weyl gauge shown in Figure \ref{fig2}.}
\end{figure}

Pictorially, the scalar field $h^a$ has the "hedgehog" form shown in Figure \ref{fig1}, with the isospin direction of $h^a$ directed along the space-time coordinate direction $x^a$. In the BPS limit ($\lambda= 0$), the profile functions solving (\ref{eq:GGode}) are analytically known \cite{bps}:
\bea
f_h(r) &=& \coth(g v r) - \frac{1}{g v r} \, , \nonumber \\
f_A(r) &=& 1 - \frac{g v r}{\sinh(g v r)} \, .
\label{eq:bps}
\eea
The corresponding action is $4 \pi v / g$.  For $\lambda > 0$, the profile functions can be computed numerically and the action is higher than in the BPS case \cite{bogomolny-marinov,kirkman-zachos}.

\subsection{Axially Symmetric Ansatz for Instantons}
\label{axial}

There is a straightforward generalization of the spherical ansatz (\ref{eq:spherical}) which has axial symmetry \cite{rebbi-rossi,kleihaus,shnir}. This ansatz is characterized by two integers, $n$ and $m$, and six profile functions, each of which depends on the polar angle $\theta$ as well as the radial coordinate $r$. In this ansatz the fields are:
\bea
h^{\rm axial}&=& \Phi_1\, \tau_r^{(n,m)}+\Phi_2\, \tau_\theta^{(n,m)}\nonumber\\
A_r^{\rm axial}&=&-\frac{1}{2 g r}\, K_1 \, \tau_\varphi^{(n)}\nonumber\\
A_\theta^{\rm axial}&=&-\frac{1}{2g}\left(1-K_2\right)\, \tau_\varphi^{(n)}\nonumber\\
A_\varphi^{\rm axial}&=&\frac{n}{2 g}\, \sin \theta \left(K_3\,  \tau_r^{(n,m)}+(1-K_4)  \tau_\theta^{(n,m)}\right)
\label{axialansatz}
\eea
Here the $\tau$ matrix combinations  $\tau_\varphi^{(n)}$, $\tau_r^{(n,m)}$ and  $\tau_\theta^{(n,m)}$ are defined as:
\bea
 \tau_\varphi^{(n)}&=& -\sin (n \varphi)\, \tau_1 +\cos(n \varphi) \, \tau_2  \nonumber\\
 \tau_r^{(n,m)}&=&  \sin (m \theta)\, \tau_\rho^{(n)} +\cos(m \theta) \, \tau_3 \nonumber\\
 \tau_\theta^{(n,m)}&=& \cos (m \theta)\, \tau_\rho^{(n)} -\sin(m \theta) \, \tau_3 \nonumber\\
  \tau_\rho^{(n)}&=&  \cos (n \varphi)\, \tau_1 +\sin(n \varphi) \, \tau_2
 \label{taus}
 \eea
The spherical ansatz (\ref{eq:spherical}) is recovered when $n=m=1$, $K_1=K_3=\Phi_2=0$, $\Phi_1=\frac{v}{2}f_h$, $K_2=K_4=1-f_A$, and $\Phi_1$, $K_2$ and $K_4$ depend only on $r$. With the axial ansatz, the instanton equations (\ref{eq:fullEqnsOfMotion}) reduce to a set of coupled equations which can be solved numerically subject to suitable boundary conditions. At the origin:
\bea
K_1(0,\theta)=K_3(0,\theta)&=&0\nonumber\\
K_2(0,\theta)=K_4(0,\theta)&=&1\nonumber\\
\sin(m \theta) \Phi_1(0,\theta)+\cos (m \theta) \Phi_2(0,\theta)&=&0\nonumber\\
\partial_r\left[\cos(m \theta) \Phi_1(0,\theta)-\sin (m \theta) \Phi_2(0,\theta)\right]_{r=0}&=&0\, .
\label{axialbcs}
\eea
At infinity,
\bea
\Phi_1&\to& 1\quad , \quad \Phi_2\to 0\nonumber\\
K_1&\to& 0\quad , \quad K_2 \to 1-m\nonumber\\
K_3&\to& \cases{\frac{\cos \theta -\cos(m\theta)}{\sin \theta}\quad, \quad m \,\, {\rm odd}\cr
\frac{1-\cos(m\theta)}{\sin\theta}\quad, \quad m \,\, {\rm even}}\nonumber\\
K_4&\to& 1-\frac{\sin(m \theta)}{\sin \theta}\, .
\label{axialinfinity}
\eea
Solving these coupled equations for the profile functions reveals a rich set of solutions \cite{kleihaus,shnir}. These solutions exist both in the BPS limit and for scalar coupling $\lambda>0$. Previously, these solutions have been studied in the context of monopoles, but here we interpret them as instanton configurations of the 3D GG model.

\begin{enumerate}
\item $m=1$ and $n>1$: axial multi-instanton solutions with instanton number $n$, having an n-fold zero of the scalar field at the origin, and axial symmetry about the $x_3$ axis. 

\item $m>1$ and $n=1$ or $n=2$ : chains of alternating instantons and anti-instantons lined up along the $x_3$ axis. For even $m$ these solutions have total instanton number 0, while for odd $m$ they have total instanton number $\pm 1$.

\item $m>1$ and $n>2$ : there appear vortex rings of zeros of the scalar field. If $m$ is even there are $\frac{m}{2}$ such vortex rings. If $m$ is odd, the number and arrangement of vortex rings depends further on whether $m=1\, ({\rm mod}\, 4)$ or $m=3\, ({\rm mod}\,  4)$. 
\end{enumerate}

It is worth commenting on several important differences between this case and instantons in 4D Yang-Mills theory. In 4D Yang-Mills theory, general multi-instanton solutions are well-known from the ADHM construction \cite{adhm,manton}. While instanton--anti-instanton solutions on ${\bf S}^4$ and ${\bf R}^4$ have been shown to exist, and some have been constructed \cite{sadun,bor}, they are not well understood. On the other hand, in the 3D GG model, instanton--anti-instanton solutions can be easily constructed  within the axial ansatz, as in case 2 above. They are saddle-points, rather than minima, of the Euclidean action. Also note that the axial 3D GG instantons of type 3 (above) are delocalized rather than pointlike. There are no such instanton configurations in 4D Yang-Mills theory, where instantons have isolated zeros. 

\section{Vacuum Structure of 3D GG Model}
\label{ggtunnel}

The classical vacuum configurations of the 3-dimensional GG model are configurations $(h, A_1, A_2)\equiv (h, \vec{A})$ that minimize the classical potential energy functional:
\bea
V[h, \vec{A}]= \int d^2 x \left[ \half \tr F_{ij}^2 + \half D_j h^a D_j h^a + \frac{\lambda}{4}\left( h^a h^a - v^2 \right)^2  \right] \, .
\label{eq:ggpotential}
\eea
We have chosen $x_3$ as the Euclidean time direction, and use Latin indices $i, j$ to denote the spatial components.
Since $V$ is a sum of non-negative terms, the minimum, $V=0$, is attained when each term vanishes locally.  The simplest such solution is the trivial configuration:  $h =- \half v \tau^3 \; , \; \vec{A} = 0$.
All vacuum configurations are gauge-equivalent to this trivial configuration, and so they may be characterized by an SU(2) group element $U$:
\bea
h^{\rm vac}&=&-\half\, v\, U \tau^3 U^\dag 
\label{eq:hvac} \\
A_j^{\rm vac}&=&\frac{i}{g} U \partial_j U^\dag \, .
\label{eq:avac}
\eea
These classical vacua may be labeled by an integer winding number in such a way that two vacua with different winding numbers are topologically distinct. This winding number may be defined in two separate ways.

First, if we impose the boundary condition that the scalar field be constant at spatial infinity, then the spatial manifold ${\bf R}^2$ is compactified to $S^2_{\rm comp}$. When restricted to a vacuum configuration, the scalar field has constant magnitude, $h^a h^a=v^2$, and so the internal space of such fields is identified with a 2-sphere, $S^2_{\rm internal}$.  Thus, $h^{\rm vac}$ defines a map from the compactified spatial manifold, $S^2_{\rm comp}$, into $S^2_{\rm internal}$, and the winding number is the degree of this map. Consider the following functional of the scalar field $h$:
\bea
W[h] &\equiv& \frac{1}{4\pi v^3}\int d^2 x\, \epsilon_{3 i j}\, \tr\left(h\, \partial_i h\, \partial_j h\right)\nonumber\\
&=&\frac{1}{8 \pi v^3} \int d^2 x\, \epsilon_{3 i j}\, \epsilon_{a b c}\, h^a \partial_i h^b \partial_j h^c \, .
\label{eq:ggw}
\eea
For a vacuum configuration this reduces to the well-known degree of the unit map $\hat{h}^{\rm vac}: S^2_{\rm comp} \mapsto S^2_{\rm internal}$. Note that for non-vacuum configurations, $W[h]$ is well-defined but is not an integer. This is completely analogous to the Chern-Simons functional $CS[\vec{A}]$ in 4D Yang-Mills theory which is defined for all gauge field configurations, but is an integer when evaluated on vacuum configurations \cite{roman,shifman}.

The above characterization of the GG vacuum configurations in terms of the functional (\ref{eq:ggw}) refers only to the scalar fields $h$. What about the gauge fields in the GG model? There is an alternative characterization of the vacuum configurations which involves both the gauge and scalar fields. To do this, define a 2-component Abelian vector field
\be
a_j = \frac{g}{v} \, \tr\left(h\, A_j\right)\, .
\label{eq:abeliana}
\ee
The magnetic flux of this field through 2-dimensional space is
\bea
\Phi[a]\equiv \frac{1}{2 \pi} \, \int d^2x\, \epsilon_{ij}\, \partial_i a_j \, .
\label{eq:abelianflux}
\eea
When evaluated on vacuum configurations
\bea
\Phi[a^{\rm vac}] = - \frac{i}{4 \pi} \int d^2 x\,  \epsilon_{ij}\, \partial_i\, \tr\left(\tau^3 U^\dag \partial_j U\right)\, ,
\label{eq:abelianvac}
\eea
which is identical to the functional $W[h^{\rm vac}]$ in (\ref{eq:ggw}), when $h^{\rm vac}$ is of the form in (\ref{eq:hvac}).  The flux functional $\Phi[a]$ in (\ref{eq:abelianflux}) is defined for all configurations, and reduces to an integer winding number for vacuum configurations. Once again, this is analogous to the Chern-Simons functional, $CS[\vec{A}]$, in 4D Yang-Mills theory.

It is important to recognize that neither $W[h]$ nor $\Phi[a]$ is gauge invariant. This must be the case, since they label topologically distinct classical vacua, which are related by a gauge transformation. 
These {\it large} gauge transformations, $U$, have the property that 
\bea
- \frac{i}{4 \pi} \int d^2 x\,  \epsilon_{ij}\, \partial_i\, \tr\left(\tau^3 U^\dag \partial_j U\right)={\rm nonzero\,\, integer}
\label{eq:large}
\eea

Two explicit examples of large gauge transformations are as follows. Each will play a role in the next section in transforming instantons into the Weyl gauge. First, consider circularly symmetric transformations in the U(1) subgroup generated by $\vec{\tau} \cdot \hat{\azAngle}$:
\be
U_\Lambda  = \exp \left[ \half i \Lambda(\rho)\, \left(\vec{\tau} \cdot \hat{\azAngle} \right)\right] \, ,
\label{eq:largeu1}
\ee
where $ \hat{\azAngle}=(-\sin \varphi, \cos \varphi, 0)$ is the azimuthal unit vector, and $\rho$ is the 2D radial distance. The vacuum configurations (\ref{eq:hvac}) and (\ref{eq:avac}) defined by this group element are
\bea
h^{\rm vac} &=& \half v \left[ \tau^3 \cos \Lambda -  \left(\vec{\tau} \cdot \hat{\rho}\right) \sin \Lambda \right] 
\nonumber\\
&=&\frac{v}{2}\left(\matrix{\cos \Lambda & -e^{-i\varphi}\,\sin \Lambda\cr
e^{i\varphi}\,\sin \Lambda &-\cos \Lambda}\right)
\, , \nonumber \\
A_j^{\rm vac} &=& \frac{1}{2g} \left[ \left(\vec{\tau} \cdot \hat{\azAngle}\right) \hat{\rho}_j\, \frac{\partial \Lambda}{\partial \rho} + \frac{\hat{\azAngle}_j}{\rho} \left\{ \tau^3 \left(\cos\Lambda-1\right) - \left(\vec{\tau}\cdot\hat{\rho}\right) \sin\Lambda \right\} \right] \, .
\label{eq:vacexample}
\eea
We require that $\Lambda(\rho)$ goes from $j_12 \pi$ at the origin (for regularity) to $j_2\pi$ at infinity (for constant scalar field), with $j_1$ and $ j_2$ being integers.  The winding is given entirely in terms of the boundary values of $\Lambda$:
\bea
W[h^{\rm vac}] &=&\frac{1}{2} \left[ -\cos \Lambda \right]^{\rho=\infty}_{\rho=0}\nonumber\\
&=& \half \left[ (-1)^{j_1} - (-1)^{j_2} \right] \nonumber\\
&=&\Phi[a^{\rm vac}]\, ,
\label{eq:largewinding1}
\eea
and takes values in $\{0, 1\}$.  So in this ansatz, we cannot probe vacuum configurations with higher winding number.  Note that a large gauge transformation in this ansatz has odd parity for $j_2$, while a small gauge transformation has even parity for $ j_2$.

A simple generalization of this ansatz is 
\be
U_\Lambda^{(n)}  = \exp \left[ \half i \Lambda(\rho)\, \tau^{(n)}_\varphi \right] \, ,
\label{eq:largeu2}
\ee
where $\Lambda(\rho)$ is as before, but the angular dependence has changed from
$(\tau \cdot \hat{\azAngle})$ in (\ref{eq:largeu1}) to
$\tau^{(n)}_\varphi$, as defined in (\ref{taus}). The case $n=1$ reduces to the previous case in (\ref{eq:largeu1}). For general $n$ the corresponding winding number becomes
\be
W[h^{\rm vac}] = \frac{n}{2} \left[ (-1)^{j_1} - (-1)^{j_2} \right]  =\Phi[a^{\rm vac}]\, ,
\label{eq:largewinding2}
\ee
which takes any integer value.

\section{Tunneling and 3D GG Instantons}

In this section we demonstrate how instantons of the 3D GG model tunnel between topologically inequivalent vacuum configurations. Since we are interested in a Hamiltonian framework to describe tunneling, we transform the given instanton solution into the Weyl ($A_3=0$) gauge. The required gauge transformation, $U$, must satisfy
\bea
A_3= \frac{i}{g}U^\dag \partial_3 U \, ,
\label{eq:weylcondition}
\eea
where $A_3$ is the third space-time component of the original instanton gauge field. 

\subsection{Weyl gauge for spherical ansatz}

First consider the spherical ansatz single instanton (\ref{eq:spherical}) from Section \ref{spherical}. This is transformed into Weyl gauge by the gauge transformation:
\bea
U_{\rm Weyl} &=& \exp \left[ \half i \Lambda_{\rm Weyl} (\rho, x_3) \,\left( \vec{\tau} \cdot \hat{\azAngle} \right)\right] \, , \nonumber \\
\Lambda_{\rm Weyl}(\rho, x_3) &=& - \rho \int_{-\infty}^{x_3} dt \frac{f_A(\sqrt{\rho^2+t^2})}{\rho^2 + t^2}
\, .
\label{eq:trans} 
\eea
The residual time-independent gauge freedom is fixed by choosing $U(\rho, x_3 \rightarrow -\infty)= {\bf 1}$.  After making this gauge transformation, the Weyl gauge single instanton is
\bea
h^{\rm Weyl} &=& \frac{v}{2} f_h(r) \left[ \cos [Q(\rho,x_3)] \tau_3 + \sin [Q (\rho,x_3)] \,\left(\vec{\tau} \cdot \hat{\rho} \right) \right] \label{eq:transmon1}  \\
&=&\frac{v}{2}f_h(r)\left(\matrix{\cos Q(\rho,x_3) & -e^{-i\varphi}\,\sin Q(\rho,x_3)\cr
e^{i\varphi}\,\sin Q(\rho,x_3) &-\cos Q(\rho,x_3)}\right)
\, , \nonumber \\
A_j^{\rm Weyl} &=& \frac{1}{2g} \left[ \hat{\rho}_j\, a_1(\rho, x_3) \,  (\vec{\tau} \cdot \hat{\azAngle})  + \frac{\hat{\azAngle}_j}{\rho} \left\{ p_1(\rho, x_3) \tau^3 + p_2(\rho, x_3) \, (\vec{\tau} \cdot \hat{\rho}) \right\} \right] \, , \nonumber
\eea
where
\bea
Q(\rho, x_3) &=& \arccos \frac{x_3}{r} - \Lambda_{\rm Weyl} \, , \nonumber \\
a_1(\rho, x_3) &=& - f_A(r) \frac{x_3}{r^2} + \partial_\rho \Lambda_{\rm Weyl} \, , \nonumber \\
p_1(\rho, x_3) &=& -1 + \left(1 - f_A(r) \frac{\rho^2}{r^2}\right) \cos \Lambda_{\rm Weyl} + f_A(r) \frac{x_3 \rho}{r^2} \sin \Lambda_{\rm Weyl} \, , \nonumber \\
p_2(\rho, x_3) &=& f_A(r) \frac{x_3 \rho}{r^2} \cos \Lambda_{\rm Weyl} - \left(1 - f_A(r) \frac{\rho^2}{r^2}\right) \sin \Lambda_{\rm Weyl} \, .
\label{eq:transmon2}
\eea

In the infinite past, $x_3 \to -\infty$, $\Lambda_{\rm Weyl}$ vanishes and the Weyl-gauge instanton corresponds to a trivial vacuum configuration :
\bea
h^{\rm Weyl} (x_3=-\infty)&=& - \half v \tau^3 \, , \nonumber \\
A_j^{\rm Weyl} (x_3=-\infty)&=& 0 \, ,
\eea
which has winding number 0.  In the infinite future, $x_3 \to + \infty$, the Weyl-gauge instanton is a winding 1 vacuum configuration, of the form in (\ref{eq:vacexample}):
\bea
h^{\rm Weyl}(x_3=+\infty)&=& \half v f_h(r) \left[ \cos [\Lambda_{\rm Weyl}(\rho,x_3=+\infty)]\, \tau_3 \right.\nonumber\\
&&\left. \hskip 2cm - \sin [\Lambda_{\rm Weyl} (\rho,x_3=+\infty)] \left(\vec{\tau} \cdot \hat{\rho} \right)\right] \, , \nonumber \\
A_j^{\rm Weyl}(x_3=+\infty) &=& \frac{1}{2g} \left[ \hat{\rho}_j\, \partial_\rho \Lambda_{\rm Weyl}(\rho, x_3=+\infty) \,  (\vec{\tau} \cdot \hat{\azAngle})  \right. \nonumber\\
&&\left. + \frac{\hat{\azAngle}_j}{\rho} \left\{\left(\cos \Lambda_{\rm Weyl}(\rho, x_3=+\infty)-1\right) \, \tau^3 \right.\right. \nonumber\\
&&\left.\left. -\sin \Lambda_{\rm Weyl}(\rho, x_3=+\infty) \, (\vec{\tau} \cdot \hat{\rho}) \right\} \right] \, .
\label{eq:transMonopole+} 
\eea
Note that $\Lambda_{\rm Weyl}(\rho, x_3=+\infty)$ goes from 0 to $-\pi$ as $\rho$ goes from 0 to $\infty$.
Thus, we see explicitly that in the Weyl gauge the configuration $(h, \vec{A})$ interpolates from a trivial vacuum configuration at $x_3=-\infty$ to a nontrivial vacuum configuration (\ref{eq:transMonopole+}) at $x_3=+\infty$. In the Weyl gauge the instanton number $1$ of the single instanton arises as the difference between the winding numbers (1 and 0, respectively) at $x_3=\pm \infty$. 

\begin{figure}
\centerline{
\includegraphics[width=8cm]{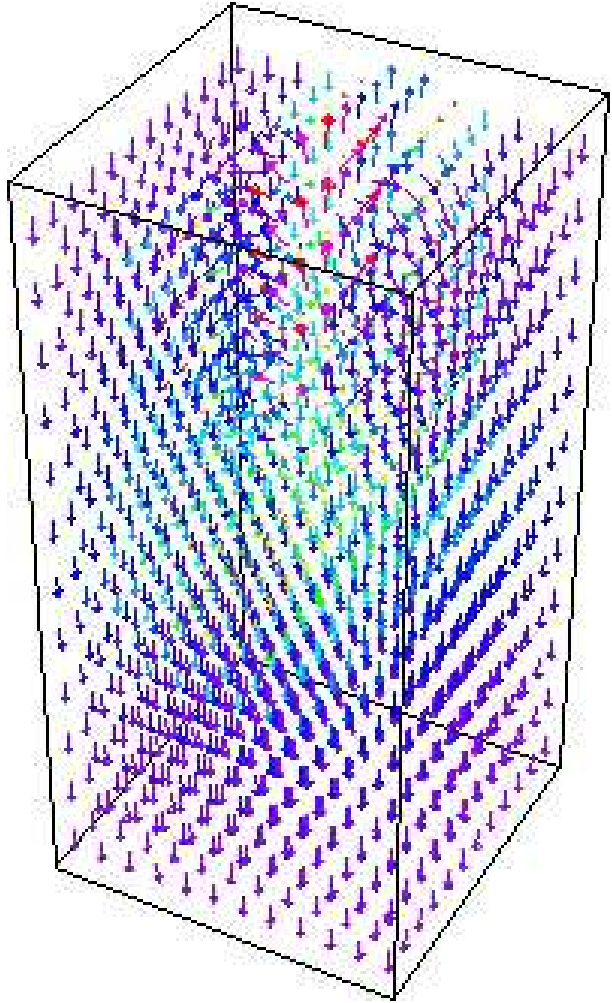}
}
\caption{\label{fig2}
The spherical ansatz single instanton scalar field $h^a$ in (\protect{\ref{eq:spherical}}) after conversion into the Weyl gauge. Contrast with the familiar hedgehog form in Figure \ref{fig1}. Except for some region
around the positive $x_3$ axis (visualized in a cross-sectional plot in Figure \ref{fig3}), the scalar field approximately takes constant vacuum value $-\frac{v}{2}\tau_3$.
}
\end{figure}

\begin{figure}
\centerline{
\includegraphics[width=8cm]{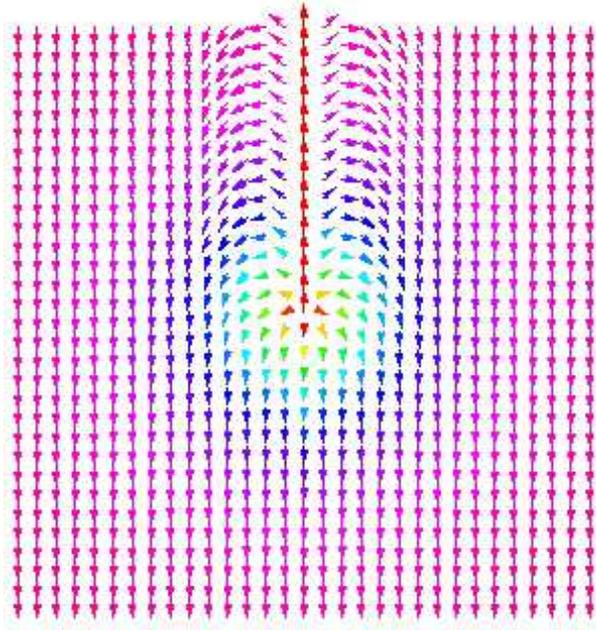}
}
\caption{\label{fig3}
A cross-section of the Weyl gauge scalar field from Figure \ref{fig2}, plotted in the $(x_1,x_3)$ plane. Notice that the scalar field has approximate value $-\frac{v}{2}\tau_3$ outside of a sheath-like region enclosing the origin and the positive $x_3$ axis. This sheath has scale set by the instanton core size scale, which in this plot is 1, and this plot covers a $10 \times 10$ square.
}
\end{figure}

\begin{figure}
\centerline{
\includegraphics[width=8cm]{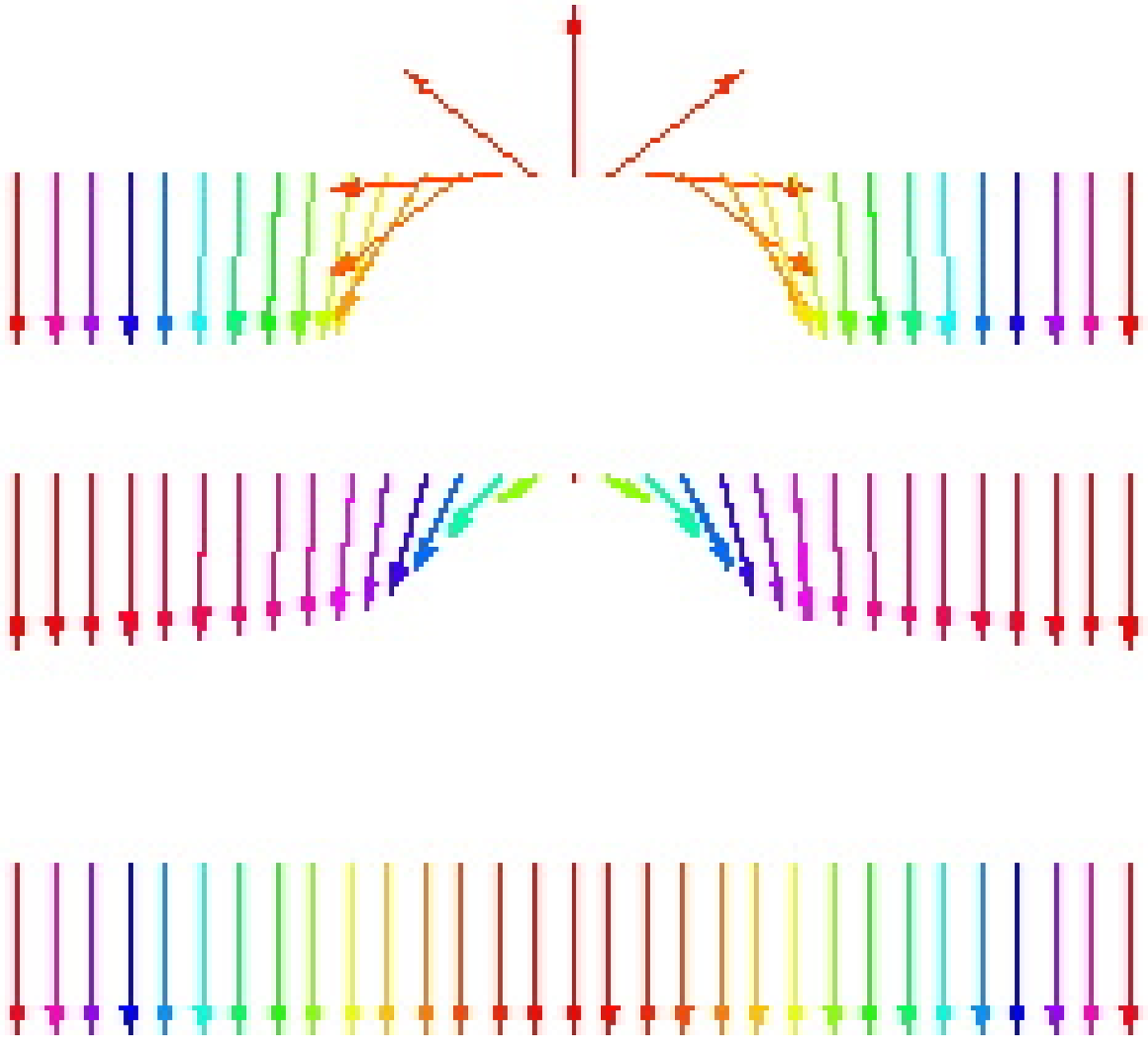}
}
\caption{\label{fig4}
Cross-sections of the Weyl gauge scalar field from Figures \ref{fig2} and \ref{fig3}, plotted along the $x_1$ axis at three different values of $x_3$: top: $x_3=20$; middle: $x_3=0$; bottom: $x_3=-20$. The bottom face approximates the trivial configuration at $x_3=-\infty$, the top face approximates a winding 1 configuration at $x_3=+\infty$, and the middle plot corresponds to a sphaleron.
}
\end{figure}

\begin{figure}
\centerline{
\includegraphics[width=10cm]{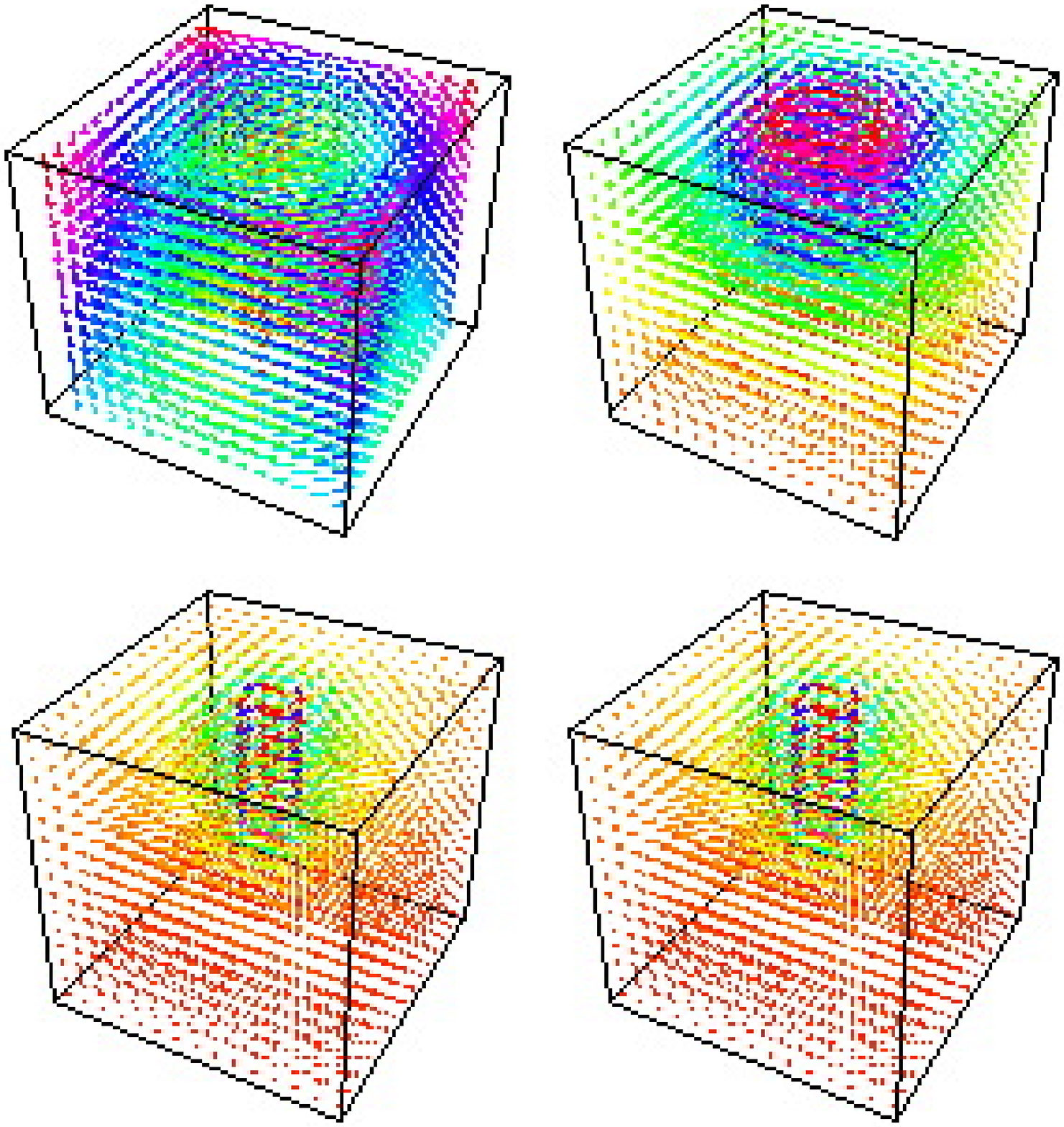}
}
\caption{\label{fig5}
The third isospin component vector field, $\vec{A}^3$, of the gauge field in the Weyl gauge, for four different values of the instanton scale parameter $L$: top left: $L=10$; top right: $L=1$; bottom left: $L=0.1$; bottom right: $L=0.01$. This isospin component, $\vec{A}^3$, of the Weyl gauge field is everywhere smooth, but notice that the circulation is concentrated around the positive $x_3$ axis, and as $L$ decreases, $\vec{A}^3$ tends to the form of the abelian Dirac monopole gauge field, as plotted in Figure \ref{fig6}.}
\end{figure}

\begin{figure}
\centerline{
\includegraphics[width=5cm]{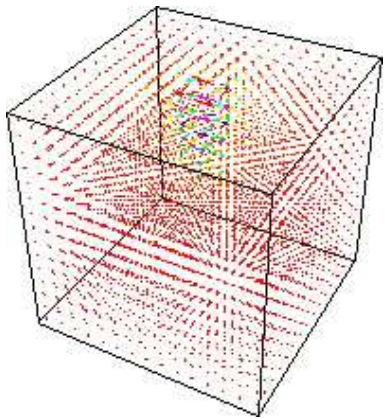}
}
\caption{\label{fig6}
The abelian vector field $\vec{A}_{\rm Dirac}=(-x_2,x_1,0)/(4\pi (r-x_3)\rho)$, which is the vector potential for a unit charge Dirac monopole, with a Dirac string singularity along the positive $x_3$ axis.
}
\end{figure}

Figures \ref{fig2} -- \ref{fig6} present visual images of the scalar and gauge configurations in the Weyl gauge. In these figures we have used the BPS forms for the profile functions $f_h$ and $f_A$, but the essential features do not change as these profile functions are varied for $\lambda > 0$.  Note first of all that these plots are very different from the familiar spherical ansatz form. For example, Figure 2 shows the scalar field, $h^a_{\rm Weyl}$, and this is clearly very different from the familiar "hedgehog" form shown in Figure \ref{fig1}. Figure 3 shows the scalar field in the $(x_1, x_3)$ plane. Note the presence of a string-like configuration along the positive $x_3$ axis. Indeed, the Weyl gauge provides a simple regularization of the singular gauge in which the scalar field $h$ has its constant vacuum value everywhere in 3D space-time. There is a sheath, of width set by the instanton core size, which surrounds the origin and extends along the positive $x_3$ axis, such that outside this sheath the scalar field has its constant vacuum value $-\frac{v}{2}\tau_3$. But inside this sheath the scalar field has core structure in such a way that it points in the positive $\tau_3$ direction on the positive $x_3$ axis. It is everywhere smooth.
Another view of the scalar field is given in Figure 4, which shows a view of $h^a$ along the $x_1$ axis at three different "times": $x_3=+\infty, 0, -\infty$. Note that on the top face the scalar field winds once in the $\theta$ direction, while at $x_3=0$ there is no net winding since $h$ vanishes at the center. Finally, on the bottom face $h$ is constant.

To visualize the gauge field we note that at $x_3=-\infty$, $A_j^{\rm Weyl}$ vanishes, while at $x_3=+\infty$, the isospin-3 component, $A_j^{3,({\rm Weyl})}$, has a vortex profile. Thus the abelian projected gauge field $a_j$ defined in (\ref{eq:abeliana}) is an abelian vortex of vorticity 1 at $x_3=+\infty$ . In the singular gauge, $A_j^{3,({\rm singular})}$ is the gauge field for an abelian Dirac monopole. In Figure 5 we plot $A_j^{3,({\rm Weyl})}$  for decreasing values of the instanton core size scale, and we see that this vector field does indeed tend to that of an abelian Dirac monopole with a Dirac string along the positive $x_3$ axis, which is shown for comparison in Figure 6. This confirms that the Weyl gauge is a regularization of the singular gauge, with the singular gauge being realized when the core size vanishes. But for nonzero core size, the Weyl gauge field $A_j^{\rm Weyl}$ is everywhere smooth.

Finally, we note that the winding 1 configuration at $x_3=+\infty$ is related to the $x_3=-\infty$ trivial vacuum configuration by a large gauge transformation:
\bea
U_{\rm large} & = &  \exp \left( \half i\, \Lambda_{\rm Weyl} (\rho, x_3 =+ \infty) \vec{\tau} \cdot \hat{\azAngle} \right) i \tau^1 \, , \nonumber \\
h^{\rm Weyl}(x_3=+\infty) &=& U_{\rm large} \, h^{\rm Weyl}(x_3=-\infty) \, U^\dag_{\rm large}\, , \nonumber \\
A_j^{\rm Weyl}(x_3=+\infty) &=& U_{\rm large} \left(A_j^{\rm Weyl}(x_3=-\infty) + \frac{i}{g} \partial_j \right) U^\dag_{\rm large} \, .
\eea
$U_{\rm large}$ is a large gauge transformation because $\Lambda_{\rm Weyl} (\rho, x_3 \rightarrow \infty)$ goes from $0$ to $(-\pi)$ as $\rho$ varies from $0$ to $\infty$.  The $i \tau^1$ part of the transformation simply flips the scalar field and is topologically trivial.  

\subsection{Weyl gauge for axial ansatz}

The axial ansatz instantons described in Section \ref{axial} may also be transformed into the Weyl gauge  in a simple manner. Note that in the axial ansatz (\ref{axialansatz}), the third space-time component of the gauge field is:
\bea
A_3=-\frac{1}{2 g r^2}\left[x_3\, K_1-\rho\,  (1-K_2)\right]\, \tau_\varphi^{(n)}
\label{axiala3}
\eea
The required gauge transformation $U$ satisfying (\ref{eq:weylcondition}) is
\bea
U^{\rm Weyl} = \exp\left(\frac{1}{2} i \Lambda_{\rm Weyl}\, \tau_\varphi^{(n)}\right)\, ,
\label{axialweylu}
\eea
where $\tau_\varphi^{(n)}$ is the generator combination defined in (\ref{taus}). The function $\Lambda_{\rm Weyl}$ is 
\bea
\Lambda_{\rm Weyl}(\rho, x_3)=\int_{-\infty}^{x_3} \frac{dt}{t^2+\rho^2} \left[t \, K_1-\rho\, (1-K_2)\right]
\eea
In the spherical case $K_1=0$, $n=1$ and $1-K_2=f_A$, so (\ref{axialweylu}) reduces to the gauge transformation (\ref{eq:trans}) discussed previously.

\section{Sphaleron of the 3D GG Model}
\label{sphaleron}

Having seen how the GG instanton describes tunneling between topologically inequivalent vacuum configurations in 2+1 dimensions, we now demonstrate the existence of a sphaleron barrier between the vacua, which is in fact the Nielsen-Olesen vortex \cite{nielsen-olesen} embedded in the Georgi-Glashow model. Any continuous interpolation between two vacuum scalar configurations with different winding numbers cannot lie completely in the vacuum manifold. There exists a configuration with maximum energy along any such interpolation, and the configuration corresponding to the minimax of these energies (if it exists), when all interpolations are considered, should be a solution to the equations of motion.  It will be a sphaleron barrier between the vacua and would correspond to a saddle point of the energy functional \cite{manton-klinkhamer,manton}.  

Consider the static configuration obtained by taking the $x_3=0$ slice of the Weyl gauge instanton in eqs.~(\ref{eq:transmon1}, \ref{eq:transmon2}), and taking the profile functions to be unspecified functions of $\rho$ that go from $0$ to $1$ as $\rho$ goes from 0 to $\infty$.  We fix the time-independent gauge freedom so that $\Lambda(\rho, x_3=0) = 0$, rather than $\Lambda(\rho, x_3=-\infty) = 0$ as was previously done. Then we obtain
\bea
h(\vec{x}) &=& \half v f_h(\rho) \vec{\tau} \cdot \hat{x} \, , \nonumber \\
A_j (\vec{x}) &=& - \half \frac{f_A(\rho)}{g \rho} \,\hat{\azAngle}_j \tau^3 \, .
\label{eq:vortexAnsatz}
\eea
By construction, this is a static configuration along an interpolation from a winding 0 vacuum to a winding 1 vacuum.  It has the form of a vortex embedded in the Georgi-Glashow model.  Plugging into the equations of motion (\eq{eq:fullEqnsOfMotion}), we get the Nielsen-Olesen equations \cite{nielsen-olesen}
\bea
\frac{d^2 f_h}{d \rho^2} + \frac{1}{\rho}\frac{d f_h}{d \rho} &=& \frac{f_h}{\rho^2}(1-f_A)^2 + \lambda v^2 f_h (f_h^2 - 1) \, , \nonumber \\
\frac{d^2 f_A}{d \rho^2} - \frac{1}{\rho} \frac{d f_A}{d \rho} &=& g^2 v^2 f_h^2 (f_A - 1) \, ,
\eea
which may be solved numerically for a given choice of the theory parameters.  Thus, the Nielsen-Olesen vortex is an embedded solution in the 3D Georgi-Glashow model, along an interpolation from a winding 0 vacuum configuration to a winding 1 vacuum configuration.  Since we expect a sphaleron solution along such an interpolation, it is suggestive that this vortex is in fact the sphaleron.  

\begin{figure}
\centerline{
\includegraphics{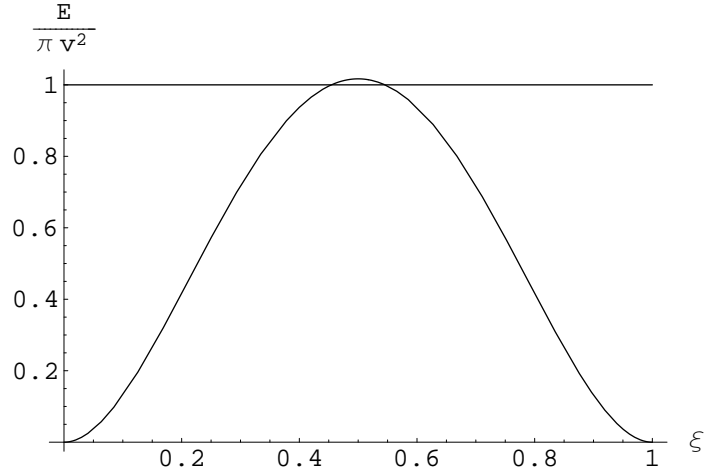}
}
\caption{\label{fig:sphaleron}
Energy as a function of $\xi$, minimized within the linear interpolation ansatz in eqs.~\ref{eq:linear1}, \ref{eq:linear2}, \ref{eq:linear3}, \ref{eq:linear4}, for the case of degenerate Higgs and $W^\pm$.  The horizontal line corresponds to the Nielsen-Olesen vortex energy.
}
\end{figure}

In order to explore this further, we consider a simple linear interpolation, parametrized by $\xi$ which goes from 0 to 1:
\bea
h (\vec{x}; \xi) &=& (1-\xi) h^{(0)} (\vec{x}) + \xi h^{(1)} (\vec{x}) \, , \nonumber \\
A_j (\vec{x}; \xi) &=& (1-\xi) A_j^{(0)} (\vec{x}) + \xi A_j^{(1)} (\vec{x}) \, , 
\label{eq:linear1}
\eea
where the winding 0 vacuum is chosen to be
\be
h^{(0)} (\vec{x}) = \half v \tau^3 \; , \; A_j^{(0)} (\vec{x}) = 0 \, ,
\label{eq:linear2} 
\ee
and the winding 1 vacuum is a large gauge transformation of the above by 
\be
U = \exp\left(\half i \Lambda(\rho) \, \vec{\tau} \cdot \hat{\azAngle} \right) i \tau^1 \, , 
\label{eq:linear3}
\ee
and where $\Lambda$ goes from 0 to $\pi$ as $\rho$ goes from 0 to $\infty$.  The energy along the interpolation is
\bea
E &=& 4 \pi v^2 \xi^2 ( 1 - \xi)^2 \int_0^\infty d  \rho \Biggl[ \frac{1}{g^2 v^2 \rho}\left(\Lambda' \sin \frac{\Lambda}{2} \right)^2 + \rho \left(\Lambda' \cos \frac{\Lambda}{2} \right)^2 +  \nonumber \\
 & & \hspace{8em} \frac{1}{\rho} \sin^2 \Lambda + 2 \lambda v^2 \rho \cos^4 \frac{\Lambda}{2}  \Biggr] \, , 
\eea
which is manifestly maximum at $\xi = 1/2$.  We choose an ansatz
\be
\Lambda = \pi \tanh (\rho/w) \, , 
\label{eq:linear4}
\ee
and minimize the energy with respect to the width parameter $w$ in order to find an upper bound on the sphaleron energy.  To facilitate comparison with the Nielsen-Olesen vortex, we choose the Higgs and the gauge masses to be degenerate which allows the Bogomolnyi bound to be saturated and gives a vortex mass of $\pi v^2$.  We find that the approximate sphaleron energy is minimum for $g v w \cong 2.5$ with a corresponding energy of $1.02 \pi v^2$.  In fig.~\ref{fig:sphaleron} we plot this minimum energy as a function of the interpolation parameter $\xi$.  When we minimize within other ansatze for $\Lambda$, we find similar energies just above the vortex mass.  Thus it seems that the Nielsen-Olesen vortex is indeed the sphaleron barrier between topologically inequivalent vacua in the 2+1 dimensional Georgi-Glashow model. This is in agreement with the work of \cite{tigran}, which contains an extensive study of the analytic and numerical properties of 3D SU(2) GG sphalerons, including sphalerons connecting vacua of higher winding number, and the magnetic and electric properties of these sphalerons.

\section{Fermions in the 3D Georgi-Glashow Model}

In order to further probe the physics associated with the instanton and the sphaleron, we add to the theory fermions in the fundamental representation of $SU(2)$. The fermion sector is defined by the action
\be
S_F = \int d^3 x \left[ \bar{\Psi} i \gamma^\mu D_\mu \Psi - f \bar{\Psi} h \Psi  \right] \, , 
\ee
where $\Psi$ is a doublet in isospace:
\be
\Psi = \left( \begin{array}{c} t \\ b \end{array} \right) \, , 
\ee
with $t, b$ being two-component spinors.  The covariant derivative is
\be
D_\mu \Psi = \partial_\mu \Psi - i g A_\mu \Psi \, .
\ee
The Yukawa coupling gives the doublet a mass 
\be
m_F = \half f v \, , 
\ee
when after spontaneous symmetry breaking the scalar field has a vacuum expectation value of magnitude $v$. Under an $SU(2)$ gauge transformation, $U$, the fermion field transforms as 
$\Psi \rightarrow U \Psi$, and the fermion action is gauge-invariant.    

The fermion action is invariant under a global phase rotation
\be
\Psi \rightarrow e^{i \alpha} \Psi \, . 
\ee
The corresponding Noether current is the fermion number current
\be
J^\mu = \bar{\Psi} \gamma^\mu \Psi \, , 
\ee
and the fermion number is conserved.  However, as we now demonstrate, this symmetry is anomalously broken by spectral flow effects.  The fermion number is changed by instanton tunneling through the sphaleron barrier (at low temperatures) and thermal transitions over the sphaleron (at high temperatures).  This is completely analogous to the electroweak theory in 3+1 dimensions.

\begin{figure}
\centerline{
\includegraphics[angle=270,width=12cm]{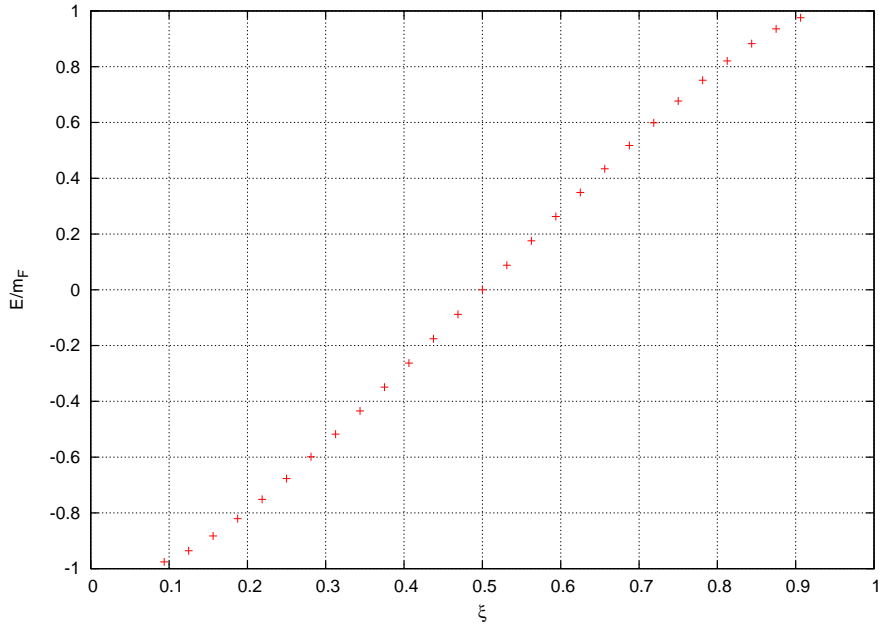}
}
\caption{\label{fig:DiracSpectrum}
The two-dimensional fermion energy spectrum of bound states, as a function of the interpolation parameter $\xi$, along the linear interpolation between winding 0 vacuum and winding 1 vacuum in (\ref{eq:linear1}, \ref{eq:linear2}, \ref{eq:linear3}, \ref{eq:linear4}).  The width, $w$, is chosen to be 1.
}
\end{figure}

The instanton can be thought of as a sequence of static configurations parametrized by the Euclidean time $x_3$.  Consider the fermion Hamiltonian in the background of the sequence of configurations (in the Weyl gauge):
\be
H_F = -i \alpha^i \left( \partial_i - i g A_i \right) + f h \gamma^0 \, ,  
\ee
where $\alpha^i = \gamma^0 \gamma^i$.  We have shown earlier that, as $x_3 \rightarrow - \infty$, the background is a winding 0 vacuum configuration, and as $x_3 \rightarrow \infty$, it is a winding 1 vacuum configuration.  So in the infinite past and the infinite future the two-dimensional Dirac energy spectrum is identical to the free spectrum (continuum of states above $m_F$ and below $-m_F$ with the Dirac sea filled in the vacuum).  At $x_3=0$, the configuration is of the form of an embedded Nielsen-Olesen vortex 
(\ref{eq:vortexAnsatz}), characterized by the profile functions $f_h(\rho)$ and $f_A(\rho)$, up to a time-independent gauge transformation.  It is well-known that the fermion has a zero-energy mode in such a background \cite{jackiw-rossi}.  In the basis $\gamma^0 = \sigma^3 , \gamma^1 = i \sigma^1 , \gamma^2 = i \sigma^2$, this zero mode is
\be
\Psi_0 = \left( \begin{array}{c} t_0 \\ b_0 \end{array}  \right) \, , \quad t_0 = f(\rho) \left( \begin{array}{c} 0 \\ 1 \end{array}  \right) \, , \quad  b_0 = f(\rho) \left( \begin{array}{c} 1 \\ 0 \end{array}  \right) \, ,
\ee  
where
\be
f(\rho) \propto \exp \left[ - \int_0^{\rho} d\rho' \left( \frac{f_A(\rho')}{2 \rho'} + m_F f_h(\rho')  \right) \right] \, .
\ee
This implies that as $x_3$ goes from $-\infty$ to $\infty$, a fermion level leaves one continuum, crosses 0 at $x_3=0$, and enters the other continuum.  This spectral flow results in the violation of fermion number by one, a la \cite{callan-dashen-gross2,kiskis}.  For example, if the direction of the crossing is from  below as $x_3$ increases, then a filled sea state becomes a filled particle state and thus a fermion is created.  In the other direction of level crossing, an empty state enters the Dirac sea and this hole corresponds to the creation of an anti-fermion. 

The level crossing persists for any background process that interpolates between topologically inequivalent vacuum configurations, and does not require an exact instanton process.  For example, consider the linear interpolation between a winding 0 vacuum and a winding 1 vacuum as described in (\ref{eq:linear1}, \ref{eq:linear2}, \ref{eq:linear3}).  This results in a level crossing from below, and thus the creation of a fermion, as shown in Figure \ref{fig:DiracSpectrum}.  We compute the bound state energies numerically, in partial waves labeled by the sum of total angular momentum and isospin ($G_z = L_z + \half \sigma^3 + \half \tau^3$), using a shooting method.  We find a single bound state in the mass gap and this has $G_z=0$.

\section{Conclusions}

We have shown how to transform instantons of the 3D Georgi-Glashow model into the Weyl ($A_3=0$) gauge. This gauge is unfamiliar in the context of such solutions viewed as 3D monopoles, but it makes explicit their interpretation as semiclassical tunneling solutions in 3D Euclidean space for the GG model. In the Weyl gauge, the instantons interpolate between classical vacuum configurations. These vacuum configurations may be labeled by two different types of winding number, one which refers only to the scalar fields and one which combines the scalar and gauge fields. The Weyl gauge also provides a simple regularization of the singular unitary gauge in which the scalar field is taken to be everywhere constant. We have also shown that the 3D GG model supports a sphaleron solution, and have shown how the spectral flow picture works when fundamental fermions are coupled to the GG fields.

\section*{Acknowledgments}

We thank Alex Kovner for discussions, and we acknowledge the support of the US DOE through grant 
DE-FG02-92ER40716. We also thank Paul Sutcliffe for bringing \cite{sadun,bor} to our attention.

\end{document}